\documentclass[reprint,amsmath,amssymb,twocolumn]{revtex4-2}

\usepackage{graphicx,amsmath,amsfonts}
\usepackage{amssymb}
\usepackage{dcolumn}
\usepackage{mathrsfs}
\usepackage{xcolor}
\usepackage{calligra}
\usepackage{aurical}
\usepackage[T1]{fontenc}
\usepackage{bm}
\usepackage{enumerate}
\usepackage{pgfplots}
\usepackage{tikz}
\usetikzlibrary{decorations.markings}
\usetikzlibrary{plotmarks}
\usepackage[export]{adjustbox}

\makeatletter
\newsavebox{\@brx}
\newcommand{\llangle}[1][]{\savebox{\@brx}{\(\m@th{#1\langle}\)}%
  \mathopen{\copy\@brx\kern-0.5\wd\@brx\usebox{\@brx}}}
\newcommand{\rrangle}[1][]{\savebox{\@brx}{\(\m@th{#1\rangle}\)}%
  \mathclose{\copy\@brx\kern-0.5\wd\@brx\usebox{\@brx}}}

\colorlet{LightGray}{gray!10}
\makeatother

\definecolor{matlab1}{HTML}{0072BD}
\definecolor{matlab2}{HTML}{D95319}
\definecolor{matlab3}{rgb}{0.4940,0.1840,0.5560}

\newlength{\mywidth}
\newlength{\myheight}
\newlength{\mydelta}
\newif\iffig
\figfalse

\begin{document}


\title{A $\delta$-free approach to quantization of transmission lines connected to lumped circuits}

\author{Carlo Forestiere}
\affiliation{Department of Electrical Engineering and Information Technology, Universit\`{a} degli Studi di Napoli Federico II, via Claudio 21,
 Napoli, 80125, Italy}
\author{Giovanni Miano}
\affiliation{ Department of Electrical Engineering and Information Technology, Universit\`{a} degli Studi di Napoli Federico II, via Claudio 21, Napoli, 80125, Italy}
\begin{abstract}
    The quantization of systems composed of transmission lines connected to lumped circuits poses significant challenges, arising from the interplay between continuous and discrete degrees of freedom. A widely adopted strategy, based on the pioneering work of Yurke and Denker, entails representing the lumped circuit contributions using Lagrangian densities that incorporate Dirac $\delta$-functions. However, this approach introduces complications, as highlighted in the recent literature, including divergent momentum densities, necessitating the use of regularization techniques. In this work, we introduce a $\delta$-free Lagrangian formulation for a transmission line coupled to a lumped circuit without the need for a discretization of the transmission line or mode expansions. This is achieved by explicitly enforcing boundary conditions at the line ends in the principle of least action. In this framework, the quantization and the derivation of the Heisenberg equations of the network are straightforward. We apply our approach to an analytically solvable network consisting of a semi-infinite transmission line capacitively coupled to a LC circuit.
\end{abstract}
\maketitle


\section{Introduction}

The transmission line paradigm plays a crucial role in the field of Circuit Quantum Electrodynamics. It is used to model one-dimensional resonators, wiring systems for control and measurement purposes, as well as losses resulting from interactions with the surrounding environment (e.g., \cite{blais_circuit_2021}). There exists an extensive body of literature on the quantization of superconducting networks, which comprise distributed circuits, transmission lines, and general impedance environments coupled with lumped circuits. Two seminal papers in this field date back to Yurke and Denker \cite{yurke_quantum_1984}, and Devoret \cite{devoret_m_h_quantum_1997}, in which they laid out the fundamental rules for quantization. These rules were later systematically extended to describe commonly used classes of superconducting quantum circuits \cite{burkard_multilevel_2004}, \cite{burkard_circuit_2005}, \cite{vool_introduction_2017}, and for quantizing black box models of electromagnetic systems \cite{nigg_black-box_2012}, \cite{solgun_multiport_2015}, \cite{minev_circuit_2021}.

Systems consisting of transmission lines coupled to lumped circuits are characterized by a complete set of commuting observables composed of both discrete and continuous field operators, and various methods have been proposed for their quantization (e.g., \cite{blais_circuit_2021}). A comprehensive summary of these methods is provided in Ref. \cite{parra-rodriguez_quantum_2018}. 
In the \textit{Yurke and Denker} approach  \cite{yurke_quantum_1984},  discrete and continuous field operators are treated on the same footing. The initial step involves the evaluation of the Lagrangian for the entire system. Following this, the canonically conjugate variables are identified, enabling the evaluation of the Hamiltonian and, subsequently, the implementation of canonical quantization. In the \textit{input-output} approach, starting from the quantization of an infinite transmission line, a semi-infinite transmission line is characterized as a one-port constituted of a resistor in series with an independent voltage source or in parallel with an independent current source (e.g., \cite{gardiner_c_w_quantum_2004}, \cite{abdo_nondegenerate_2013}, \cite{vool_introduction_2017}).  In the \textit{discrete approach}, the transmission line is initially represented as a cascade of discrete lumped elements, i.e., inductors and capacitors. Subsequently, the Lagrangian and Hamiltonian of the system are evaluated, canonical quantization is performed, and eventually, the continuum limit is taken (e.g. \cite{johansson_readout_2006}, \cite{peropadre_scattering_2013}, \cite{bamba_recipe_2014}, \cite{malekakhlagh_origin_2016}). In the \textit{mode expansion} approach, the Lagrangian of the overall system is expressed by expanding the field operators into a well-suited set of modes. Subsequently, the Hamiltonian is computed, and canonical quantization is applied (e.g., \cite{blais_circuit_2021}, \cite{parra-rodriguez_quantum_2018}, \cite{bourassa_ultrastrong_2009}). 
Modes can be selected in different ways: for example, as the eigenmodes of the complete linearized system \cite{bourassa_ultrastrong_2009} or as the eigenmodes of the transmission line. In the latter case, specific boundary conditions are imposed to guarantee the absence of mode-mode coupling in the contribution of the transmission line to the Hamiltonian \cite{parra-rodriguez_quantum_2018}. In the \textit{multiport impedance} approach, a transmission line is described as a two-port, then an equivalent lumped element two-port is evaluated using classical synthesis techniques of circuit theory, and the resulting circuit is subsequently quantized (e.g., \cite{nigg_black-box_2012}, \cite{solgun_multiport_2015}, \cite{minev_circuit_2021}). 

The approach proposed by Yurke and Denker offers an interesting perspective by treating discrete and continuous degrees of freedom on equal footing, all while avoiding the need to discretize the continuous degrees of freedom or to expand them into modes. The introduction of a Lagrangian density that contains both regular functions and distributions allows handling the simultaneous presence of continuous and discrete degrees of freedom \cite{yurke_quantum_1984}. 
For example, the Lagrangian density that describes the coupling of a semi-infinite transmission line with a lumped circuit is $\mathcal{L}=u(x)\mathcal{L}_{tml}+\delta(x)L_{lump}$, where $\mathcal{L}_{tml}$ represents the Lagrangian density of the transmission line assumed to be infinite, $u(x)$ is the Heaviside function, $L_{lump}$ contains the contribution of the lumped circuit, and $\delta(x)$ is a Dirac $\delta$-function located at the end of the line. As recently highlighted in \cite{parra-rodriguez_canonical_2022}, this approach may lead to conjugate momentum densities that involve Dirac $\delta$- functions, for instance when the lumped circuit is capacitively coupled to the flux variable of the line or inductively coupled to the charge variable. Consequently, there may be the need to handle products of distributions, which may require regularization procedures. 

In this paper, we revisit the Yurke and Denker approach by introducing a Lagrangian formulation that eliminates the need for distributions. We achieve this by introducing a Lagrangian formulation that explicitly enforces the boundary conditions at the ends of the transmission line in the principle of least action. The resulting Euler-Lagrange equations, Hamilton equations and Heisenberg equations validate the physical soundness of our approach. To simplify the presentation, we focus on the common scenario where a semi-infinite transmission line is connected to a lumped circuit via a single capacitor. 
 The formulation we have established seamlessly extends to more complex scenarios. For example, if the transmission line were coupled to the lumped circuit via both inductive and capacitive elements, we would simply incorporate the contribution of the inductors into the interaction term \ref{eq:Lint}. In this situation, the system of Euler-Lagrange equations would additionally encompass the characteristic equation associated with the coupling inductor. Extending this formulation to finite length transmission lines and multiconductor transmission lines is also straightforward.

 In Section II, we focus on the Lagrangian and Hamiltonian formulations for a semi-infinite transmission line capacitively coupled to a lumped circuit. In Section III, we give the quantum model and derive the Heisenberg equations of motion for the conjugate observables of the entire system. In Section IV, we obtain a reduced quantum model of the lumped circuit. In Section V, we apply the proposed approach to an analytically solvable network. In Section VI, we conclude with a discussion of the main achievements.

\section{Semi-infinite transmission line capacitively coupled to a lumped circuit}

\label{sec:SemiInf}
\begin{figure}
    \centering
    \includegraphics[width=\columnwidth]{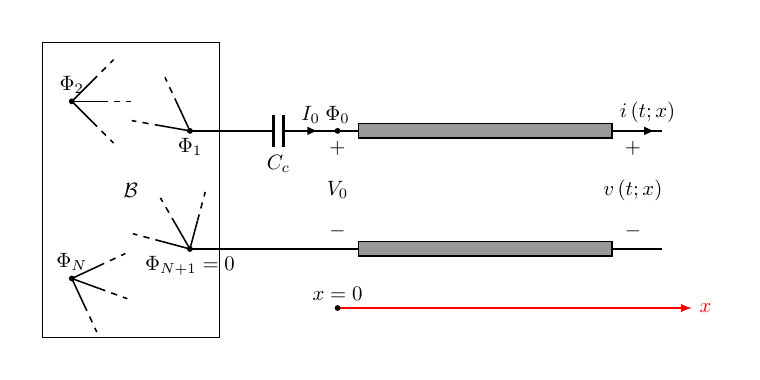}
    \caption{A network composed of a semi-finite transmission line coupled to a lumped circuit $\mathcal{B}$ through the linear capacitor $C_c$.}
    \label{fig:SemiInfinite_TL_Network}
\end{figure}

We consider a semi-infinite transmission line connected to a lumped circuit $\mathcal{B}$ through the linear capacitor $C_c$, as shown in Fig. \ref{fig:SemiInfinite_TL_Network}. The lumped circuit $\mathcal{B}$ contains linear capacitors, linear inductors, and Josephson junctions. All nodes of $\mathcal{B}$ are active, that is, in any node at least one capacitor and one inductor, either linear or Josephson junction, met. This system represents a common linear coupling configuration in circuit QED (e.g., \cite{blais_circuit_2021}). 

 The entire network has $(N+2)$ nodes, labeled with integers from $0$ to $N+1$, as illustrated in Fig. \ref{fig:SemiInfinite_TL_Network}. The coupling capacitor $C_c$ is connected to the nodes $0$ and $1$, and the transmission line is connected to the nodes $0$ and $N+1$. We employ the node method (e.g., \cite{vool_introduction_2017}) to describe the part of the system made of lumped elements. We indicate the node fluxes by $\Phi_0(t), \Phi_1(t), ..., \Phi_N(t), \Phi_{N+1}(t)$ where $\Phi_{N+1}=0$. The voltage of the branch $h$ connected to the nodes $i$ and $j$ is $(\dot\Phi_i - \dot\Phi_j)$, where the dot denotes the ordinary time derivative.

The fundamental electrical variables of the semi-infinite transmission line are the current intensity $i(t;x)$ and the voltage $v(t;x)$, where $0 \le x<\infty$. We denote by $\ell$ the inductance per unit length and by $c$ the capacitance per unit length of the line. We introduce the field $\phi(t;x)=\int_{-\infty}^t v(\tau;x)d\tau$, thus $v=\phi_t$ and $i=-\phi_x/\ell$ where $\phi_t\equiv\partial\phi/\partial t$ and $\phi_x\equiv \partial\phi/\partial x$. 

We describe the capacitive coupling between the transmission line and the lumped circuit by  explicitly imposing  the boundary conditions
\begin{subequations}
\begin{align}
\label{eq:voltage0}
    v(t;x=0)&=V_0(t), \\
    \label{eq:current1}
    i(t;x=0)&=I_0(t),
\end{align}
\end{subequations}
where $I_0(t)$ is the current intensity through the coupling capacitor $C_c$ and $V_0(t)$ is the voltage at the end $x=0$ of the line, according to the reference current direction shown in Fig. \ref{fig:SemiInfinite_TL_Network}. Since $V_0=\dot\Phi_0$, in our approach the boundary condition \ref{eq:voltage0} is satisfied by imposing 
\begin{equation}
\label{eq:Phi1}
    \phi(t;x=0)=\Phi_0(t).
\end{equation}
The other boundary condition will be naturally imposed through the Lagrangian formulation of the problem.

\subsection{Lagrangian and Euler-Lagrange equations}

The Lagrangian $L$ for the entire system consists of three distinct components: the contribution ${L}_b$ arising from the one-port $\mathcal{B}$, the contribution ${L}_{tml}$ originating from the transmission line, and the contribution ${L}_{int}$ stemming from the capacitive coupling between $\mathcal{B}$ and the transmission line through the capacitor $C_c$,
\begin{equation}
\label{eq:Lagr}
{L}={L}_b+{L}_{{tml}}+{L}_{int}.
\end{equation}

The variables $\Phi_1, \Phi_2, ..., \Phi_N$ are the degrees of freedom of $\mathcal{B}$. We define the $N$- dimensional column vector $\bold{\Phi} = |\Phi_1, \Phi_2, ..., \Phi_N |^\intercal $.  Additionally, we denote with $C_{rs}$ the capacitance of the capacitor linking nodes $r$ and $s$ with $r,s=1,2, ..., N+1$ (where $C_{rs}$ equals 0 in cases where no capacitor connects these nodes). We first introduce the $(N+1)\times (N+1)$ capacitance matrix $[\mathbf{C}]$, whose non-diagonal elements are $-C_{rs}$ and the diagonal elements are equal to the opposite of the sum of values in the corresponding row or column. Subsequently, we introduce the matrix $[\mathbf{C}_b]$, which is derived from $[\mathbf{C}]$ by excluding both the row and column corresponding to the ground node, as described for instance in \cite{vool_introduction_2017}. The contribution ${L}_b$ to the Lagrangian is given by
\begin{equation}
\label{eq:Lb}
{L}_b=\frac{1}{2}\dot{\bold{\Phi}}^\intercal[\bold{C}_b]\dot{\bold{\Phi}}-{U}_b\left(\bold{\Phi}\right),
\end{equation}
where ${U}_b={U}_b\left(\bold{\Phi}\right)$ is the potential energy of the circuit comprising the sum of energies stored in the inductors, whether linear or Josephson junctions (cf., \cite{burkard_multilevel_2004}, \cite{vool_introduction_2017}).
This expression would coincide with the Lagrangian of $\mathcal{B}$ if it were isolated, meaning $I_0=0$.

The degrees of freedom of the line are represented by $\Phi_0(t)$ and the field $\phi(t;x)$ for $ 0<x<\infty$. The contribution ${L}_{tml}$ is given by (e.g., \cite{yurke_quantum_1984})
\begin{equation}
\label{eq:Ltml}
{L}_{{tml}}=\int_{0}^{\infty}dx\,\mathcal{L}_{tml}\left({\phi}_t, \phi_x\right),
\end{equation}
where
\begin{equation}
\label{eq:Ldens}
\mathcal{L}_{tml }=\frac{c}{2}{\phi}_t^2-\frac{1}{2 \ell}\phi_x^2
\end{equation}
is the Lagrangian density. This expression would coincide with the Lagrangian of the semi-infinite transmission line if it were isolated, that is, $I_0=0$.
The contribution ${L}_{int}$ to the Lagrangian is given by
\begin{equation}
\label{eq:Lint}
{L}_{int}=\frac{C_{c}}{2}\left(\dot{\Phi}_1-\dot{\Phi}_0\right)^2.
\end{equation}

We now derive the Euler-Lagrange equations of the system. According to the principle of least action, we consider a generic time interval $(t_1,t_2)$ and variations of degrees of freedom equal to zero for $t=t_1$ and $t=t_2$. The degrees of freedom of the entire system are $\boldsymbol{\Phi}(t)$, $\Phi_0(t)$ and $\phi(t;x)$ for $ 0<x<\infty$.  The first variation of the action $S=\int_{t_1}^{t_2} L dt$ has three contributions, $\delta S= \delta{S}_{\mathbf{\Phi}}+\delta{S}_{\Phi_0}+\delta{S}_{\phi}$,
where 
\begin{equation}
\delta{S}_{\mathbf{\Phi}}= -\int_{t_1}^{t_2}dt \, \delta \mathbf{\Phi}^\intercal\left[\left(\frac{d}{d t}\frac{\partial {L}_b}{\partial \dot{\mathbf{\Phi}}}-\frac{\partial {L}_b}{\partial \mathbf{\Phi}}\right)+\frac{d}{d t}\frac{\partial {L}_{int}}{\partial \dot{\mathbf{\Phi}}}\right]
\end{equation}
originates from $\delta \boldsymbol{\Phi}$,
\begin{equation}
\delta{S}_{\Phi_0}= -\int_{t_1}^{t_2}dt\, \delta \Phi_0 \left(\frac{d}{d t}\frac{\partial {L}_{int}}{\partial \dot{\Phi}_0}+\left.\frac{\partial \mathcal{L}_{tml}}{\partial \phi_x}\right|_{x=0}\right)
\label{eq:deltaS_phi0}
\end{equation}
originates from $\delta {\Phi}_0$, and
\begin{equation}
\delta{S}_{\phi}= -\int_{t_1}^{t_2} dt \int_{0}^{\infty} dx\, \delta \phi\left(\frac{\partial}{\partial t}\frac{\partial\mathcal{L}_{tml}}{\partial {\phi}_t}+\frac{\partial}{\partial x}\frac{\partial \mathcal{L}_{tml}}{\partial \phi_x}\right)
\end{equation}
originates from $\delta \phi$ in $(0,\infty)$ (we remember that $\delta \Phi_0 = \delta \phi(t;x=0)$).  The degree of freedom $\Phi_0(t)$ is shared by the coupling capacitor and the transmission line, and it contributes to the variation of the action through both $L_{int}$ and $L_{tml}$. Indeed, the second term in expression \ref{eq:deltaS_phi0} is due to $L_{tml}$ and it has been derived through integration by parts, which gives rise to the boundary term at $x=0$. By enforcing a zero variation of the action due to $\delta \boldsymbol{\Phi}$, keeping $\delta \Phi_0$ and $\delta \phi$ equal to zero, we obtain the Euler-Lagrange equation for $\boldsymbol{\Phi}$,
\begin{equation}
\label{eq:eqlagrPhicir}
\frac{d}{d t}\frac{\partial}{\partial \dot{\mathbf{\Phi}}}({L}_b+L_{int})-\frac{\partial {L}_b}{\partial \mathbf{\Phi}}=0.
\end{equation}
By enforcing a zero variation of the action due to $\delta \phi$, keeping $\delta \Phi_0 =0$ and taking into account Eq. \ref{eq:eqlagrPhicir} we obtain the Euler-Lagrange equation for $\phi$ with $0< x< \infty$,
\begin{equation}
\label{eq:eqlagrphitml}
\frac{\partial}{\partial t}\frac{\partial\mathcal{L}_{tml}}{\partial {\phi}_t}+\frac{\partial}{\partial x}\frac{\partial \mathcal{L}_{tml}}{\partial \phi_x}=0.
\end{equation}
Eventually, taking into account $\ref{eq:eqlagrPhicir}$ and $\ref{eq:eqlagrphitml}$, and by enforcing a zero variation of the action due to $\delta \Phi_0$ we obtain the Euler-Lagrange equation for $\Phi_0$,
\begin{equation}
\label{eq:eqcoupl}
\frac{d}{d t}\frac{\partial {L}_{int}}{\partial \dot{\Phi}_0}+\left.\frac{\partial \mathcal{L}_{tml}}{\partial \phi_x}\right|_{x=0}=0.
\end{equation}
We highlight that $\left.\frac{\partial \mathcal{L}_{tml}}{\partial \phi_x}\right|_{x=0} \delta \Phi_0$ gives the contribution of $\delta \Phi_0$ to the variation of the Lagrangian.

From Eq. \ref{eq:eqlagrPhicir} we obtain
\begin{equation}
\label{eq:eqlagrPhicir1}
[\mathbf{C}_b]\frac{d^2\mathbf{\Phi}}{d t^2} +\frac{\partial U_b}{\partial \mathbf{\Phi}}+C_{c}\frac{d^2}{d t^2}\left({\Phi}_1-{\Phi}_0\right) \mathbf{1}_N=\mathbf{0},
\end{equation}
where $\mathbf{1}_N$ is the N-dimensional column vector $|1,0,...,0,0|^\intercal$. 
From Eq. \ref{eq:eqcoupl} we obtain
\begin{equation}
\label{eq:eqcoupl1}
C_{c}\frac{d^2}{dt^2}\left({\Phi}_1-{\Phi}_0\right)+\frac{1}{\ell}\phi_x(t;x=0)=0.
\end{equation}
From Eq. \ref{eq:eqlagrphitml} we obtain the wave equation for the flux field, for $0<x<\infty$,
\begin{equation}
\label{eq:eqlagrphitml1}
\frac{\partial^2 \phi}{\partial t^2}-v_p^2\frac{\partial^2 \phi}{\partial x^2}=0
\end{equation}
where $v_p=1/\sqrt{\ell c}$. 
We now analyze the physical meaning of Eqs. \ref{eq:eqlagrPhicir1} and \ref{eq:eqcoupl1}.
The system of Eqs. \ref{eq:eqlagrPhicir1} returns the Kirchhoff current law at the nodes $1, 2, ..., N$ of the lumped circuit $\mathcal{B}$. The characteristic equation of the capacitor $C_c$ is
\begin{equation}
\label{eq:Charact_c}
C_{c}\frac{d^2}{dt^2}\left({\Phi}_1-{\Phi}_0\right)=I_0.
\end{equation}
Since $i(t;x=0)=-\phi_x(t;x=0)/\ell$, combining Eqs. \ref{eq:eqcoupl1} and \ref{eq:Charact_c}, we obtain  $i(t;x=0)=I_0(t)$, that is, the boundary condition \ref{eq:current1}. In our approach, the boundary condition for the current intensity at the line end ($x=0$) arises as a natural boundary condition of the Lagrangian formulation, while the boundary condition for the flux \ref{eq:Phi1} has to be imposed explicitly.

Equations \ref{eq:eqlagrPhicir1}-\ref{eq:eqlagrphitml1} 
coincide with those obtained by initially modeling the semi-infinite transmission line as a series of discrete LC lumped elements with length $\Delta x$. Subsequently, we evaluate the Lagrangian of this discrete cascade, establish the Euler-Lagrange equations, and ultimately approach the continuum limit as $\Delta x \rightarrow 0$ \cite{peropadre_scattering_2013}. 

In the discrete cascade, the independent degrees of freedom of the line are  $\Phi_0(t)=\phi(t;x=0), \phi(t;x=\Delta x), \phi(t;x=2\Delta x), \ldots$.  This agrees with the fact that, in the Lagrangian formulation of the continuum model, $\Phi_0(t)$ and $\phi(t;x)$ for $ 0<x<\infty$ can be considered as independent degrees of freedom.

\subsection{Hamiltonian and Hamilton equations}

We now introduce the conjugate variables according to the Euler-Lagrange equations \ref{eq:eqlagrPhicir}-\ref{eq:eqcoupl}. The conjugate momentum to $\mathbf{\Phi}$ is
\begin{equation}
\label{eq:Q}
\mathbf{Q}=\frac{\partial}{\partial \dot{\mathbf{\Phi}}}({L}_b+L_{int})=[\mathbf{C}_b]\dot{\mathbf{\Phi}}+C_{c}\left(\dot{\Phi}_1-\dot{\Phi}_0\right) \mathbf{1}_N.
\end{equation}
We observe that
$[\bold{C}_b]\dot{\bold{\Phi}}$ would be the conjugate momentum to  $\mathbf{\Phi}$ if the lumped circuit were isolated. The conjugate momentum to $\Phi_0$ is
\begin{equation}
\label{eq:q}
Q_0=\frac{\partial {L}_{int}}{\partial \dot{\Phi}_0}=-C_{c}\left(\dot{\Phi}_1-\dot{\Phi}_0\right).
\end{equation}
We observe that $\dot{Q_0}=-I_0$. 
Eventually, the conjugate momenta to $\phi$ is for $0<x<\infty$
\begin{equation}
\label{eq:qtml}
q=\frac{\partial \mathcal{L}_{tml}}{\partial {\phi}_t}=c{\phi}_t.
\end{equation}
We observe that $Q_0\not=q(t;x=0)$. Combining Eqs. \ref{eq:Q} and \ref{eq:q} we obtain
\begin{equation}
\dot{\mathbf{\Phi}}=[\mathbf{C}_b]^{-1}\left(\mathbf{Q}+Q_0 \mathbf{1}_N \right),
\end{equation}
and
\begin{equation}
\dot{\Phi}_0=\mathbf{p}^\intercal \mathbf{Q} +\frac{1}{C_{p}}Q_0,
\end{equation}
where $\mathbf{p}$ is the $N$ dimensional column vector with elements $p_j=[\mathbf{C}_b^{-1}]_{1,\,j}$ for $j=1,2,...,N$ (i.e., the elements of the first row of $[\mathbf{C}_b^{-1}]$) and
\begin{equation}
\frac{1}{C_p}=\frac{1}{C_c}+p_1.
\end{equation}
Since the matrix $[\mathbf{C}_b]$ is symmetric, we have $[\bold{C}_b]\mathbf{p}=\mathbf{1}_N$. 

The Hamiltonian of the system is given by
\begin{equation}
\label{eq:hamilt}
{H}=\mathbf{Q}^\intercal\dot{\mathbf{\Phi}}+Q_0\dot{\Phi}_0+\int_{0}^{\infty}q{\phi}_tdx-{L}.
\end{equation}
It is equal to the sum of the energies stored within the circuit $\mathcal{B}$, the coupling capacitor, and the transmission line. In terms of the degrees of freedom and their associated conjugate momenta, \ref{eq:hamilt} can be expressed as:
\begin{equation}
\label{eq:Ham0}
{H}={H}_b+{H}_{{tml }}+{H}_{int},
\end{equation}
where 
\begin{subequations}
\begin{align}
\label{eq:Hb}
{H}_b&=\frac{1}{2}{\bold{Q}}^\intercal[\bold{C}_b]^{-1}{\bold{Q}}+{U}_c\left(\bold{\Phi}\right), \\
\label{eq:Hint}
{H}_{int}&=\mathbf{p}^\intercal \mathbf{Q}Q_0+\frac{1}{2C_p}Q_0^2,
\\ 
\label{eq:Html}
{H}_{{tml}}&=\int_{0}^{\infty}\left(\frac{1}{2c} {q}^2+\frac{1}{2 \ell}\phi_x^2\right) dx.
\end{align}
\end{subequations}

{
Using the Euler-Lagrangian equations and the the expressions of the conjugate momenta, we obtain for the total differential of the Hamiltonian \ref{eq:hamilt}: 
\begin{multline}
 \label{eq:dH2}
{dH}= \dot{\mathbf{\Phi}}^\intercal d\bold{Q}-\dot{\bold{Q}}^\intercal d\mathbf{\Phi}+
\dot{\Phi}_0 d{Q}_0 -\dot{Q}_0 d{\Phi_0}
\\
+\int_0^\infty \left[\dot{\phi} dq(x) -\dot{q} d\phi(x) \right]dx. 
\end{multline}
 On the other hand, from the Hamiltonian expression \ref{eq:Ham0} we also obtain
\begin{multline}
 \label{eq:dH1}
{dH}=\left(\frac{\partial H}{\partial \bold{Q}}\right)^\intercal d\bold{Q}+\left(\frac{\partial H}{\partial \mathbf{\Phi}}\right)^\intercal d\mathbf{\Phi}+\frac{\partial H_{}}{\partial Q_0} d{Q_0}+
\frac{\partial H_{}}{\partial\Phi_0} d{\Phi}_0\\
+\int_0^\infty \left[\frac{\delta H}{\delta q(x)} dq(x)+\frac{\delta H}{\delta \phi(x)} d\phi(x)\right]dx,   
\end{multline}
where
\begin{subequations}
\begin{align}
\frac{\partial H}{\partial \bold{Q}}&=[\bold{C}_b]^{-1}{{\bold{Q}}}+\mathbf{p}{Q}_0, \\
\frac{\partial H}{\partial \bold{\Phi}}&=\frac{\partial{U}_b}{\partial \hat{{\mathbf{\Phi}}}},
\end{align}
\end{subequations}
\begin{subequations}
\begin{align}
\frac{\partial H}{\partial {Q}_0} &=\mathbf{p}^\intercal {\mathbf{Q}}+\frac{1}{C_p}{Q}_0, \\
\label{eq:Heis_6}
\frac{\partial H}{\partial {\Phi}_0}&=-\frac{1}{\ell} {\phi}_x(t;x=0),
\end{align}
\end{subequations}
\begin{subequations}
\begin{align}
\frac{\delta H}{\delta {q}(x)}&=\frac{1}{c}{q}, \\
\frac{\delta H}{\delta {\phi}(x)}&=-\frac{1}{\ell} {\phi}_{xx},
\end{align}
\end{subequations}
and ${\phi}_{xx}\equiv \partial^2 {\phi}/\partial x^2$.
In particular, the term proportional to $d\Phi_0$ arises from $L_{tml}$ as a boundary contribution at $x=0$. The Hamilton equations are obtained by imposing that the expressions \ref{eq:dH2} and \ref{eq:dH1}  are equal for every $d\bold{Q}$, $d\mathbf{\Phi}$, $d{Q}_0$, $d{\Phi_0}$, $dq(x)$ and $d\phi(x)$.

The Hamilton equations for the conjugate variables (${\mathbf{\Phi}}, {\mathbf{Q}}$) are:
\begin{subequations}
\begin{align}
\label{eq:Ham1}
\dot{\hat{\mathbf{\Phi}}}&=\frac{\partial H}{\partial \bold{Q}}=[\bold{C}_b]^{-1}{{\bold{Q}}}+\mathbf{p}{Q}_0, \\
\label{eq:Ham2}
\dot{{\mathbf{Q}}}&=-\frac{\partial H}{\partial \bold{\Phi}}=-\frac{\partial{U}_b}{\partial \hat{{\mathbf{\Phi}}}},
\end{align}
\end{subequations}
where $\frac{\partial H}{\partial {Q}_i}$ and $-\frac{\partial H}{\partial {\Phi}_i}$ coincide with the Poisson brackets $\{\Phi_i,H\}$ and $\{Q_i,H\}$, respectively. The Hamilton equations for the conjugate variables ($\Phi_0,Q_0$) are:
\begin{subequations}
\begin{align}
\label{eq:Ham3}
\dot{{\Phi}}_0&=\frac{\partial H}{\partial {Q}_0} =\mathbf{p}^\intercal {\mathbf{Q}}+\frac{1}{C_p}{Q}_0, \\
\label{eq:Ham4}
\dot{{Q}}_0&=-\frac{\partial H}{\partial {\Phi}_0}=\frac{1}{\ell} {\phi}_x(t;x=0),
\end{align}
\end{subequations}
where $\frac{\partial H}{\partial {Q}_0}$ and $-\frac{\partial H}{\partial {\Phi}_0}$ coincide with the Poisson brackets $\{\Phi_0,H\}$ and $\{Q_0,H\}$, respectively. Lastly, the Hamilton equations for the conjugate variables $(\phi, q)$ for $0<x<\infty$ are:
\begin{subequations}
\begin{align}
\label{eq:Ham5}
{\phi}_t=\frac{\delta H}{\delta {q}(x)}&=\frac{1}{c}{q}, \\
\label{eq:Ham6}
{q}_t=-\frac{\delta H}{\delta {\phi}(x)}&=\frac{1}{\ell} {\phi}_{xx},
\end{align}
\end{subequations}
where $\frac{\delta H}{\delta {Q}(x)}$ and $-\frac{\delta H}{\delta{\Phi}(x)}$ coincide with the Poisson brackets $\{\phi,H\}$ and $\{q,H\}$, respectively. Combining the Hamilton equations, we obtain the system of equations \ref{eq:eqlagrPhicir1}-\ref{eq:eqlagrphitml1}. }

The conjugate momenta, the Poisson brackets, and the Hamiltonian equations coincide with those derived when the semi-infinite transmission line is modelled as a series of discrete LC lumped elements and the continuum limit is considered. \cite{peropadre_scattering_2013}.

\section{Heisenberg equations of motion}

We now promote the conjugate variables $(\mathbf{\Phi}, \mathbf{Q})$, $({\Phi_0}, {Q_0})$, $({\phi}, {q})$ and the Hamiltonian $H$ to operators. In the Heisenberg picture, the equal-time commutation relations are
\begin{subequations}
\begin{align}
\label{eq:com1}
\left[\hat{\Phi}_k(t),\hat{Q}_k(t)\right]&=i\hbar \qquad \text{for} \; k=1,2,...,N,
\\
\label{eq:com2}
\left[\hat{\Phi}_0(t),\hat{Q}_0(t)\right]&=i\hbar,
\\
\label{eq:com3}
\left[\hat{\phi}(t;x'),\hat{q}(t;x)\right]&=i\hbar\delta(x'-x) \quad \text{for} \; 0<x<\infty,
\end{align}
\end{subequations}
while all remaining equal-time commutators vanish. 

We can now derive the Heisenberg equations for the fundamental operators of the system. For the conjugate operators ($\hat{\mathbf{\Phi}}, \hat{\mathbf{Q}}$), we have:
\begin{subequations}
\begin{align}
\label{eq:Heis_1}
\dot{\hat{\mathbf{\Phi}}}&=\frac{1}{i\hbar}\left[\hat{\mathbf{\Phi}},\hat{H}\right]=[\bold{C}_b]^{-1}{\hat{\bold{Q}}}+\mathbf{p}\hat{Q}_0, \\
\label{eq:Heis_2}
\dot{\hat{\mathbf{Q}}}&=\frac{1}{i\hbar}\left[\hat{\mathbf{Q}},\hat{H}\right]=-\frac{\partial{U}_b}{\partial \hat{{\mathbf{\Phi}}}}.
\end{align}
\end{subequations}
For the conjugate operators ($\hat\Phi_0,\hat Q_0$), we have:
\begin{subequations}
\begin{align}
\label{eq:Heis_5}
\dot{\hat{\Phi}}_0&=\frac{1}{i\hbar}\left[\hat{\Phi}_0,\hat{H}\right]=\mathbf{p}^\intercal \hat{\mathbf{Q}}+\frac{1}{C_p}\hat{Q}_0, \\
\label{eq:Heis_6}
\dot{\hat{Q}}_0&=\frac{1}{i\hbar}\left[\hat{Q}_0,\hat{H}\right]=\frac{1}{\ell} \hat{\phi}_x(t;x=0).
\end{align}
\end{subequations}
Lastly, for the conjugate operators $(\hat\phi,\hat q)$ for $0<x<\infty$, the Heisenberg equations are given by:
\begin{subequations}
\begin{align}
\label{eq:tlphi}
\hat{\phi}_t=\frac{1}{i\hbar}\left[\hat{\phi},\hat{H}\right]&=\frac{1}{c}\hat{q}, \\
\label{eq:tlq}
\hat{q}_t=\frac{1}{i\hbar}\left[\hat{q},\hat{H}\right]&=\frac{1}{\ell} \hat{\phi}_{xx},
\end{align}
\end{subequations}
where $\hat{\phi}_{xx}\equiv \partial^2\hat{\phi}/\partial x^2$. In the Appendix \ref{sec:Heisen}, we evaluate the commutators $\left[\hat{\phi},\hat{{H}}_{tml}\right]$,  $\left[\hat{q},\hat{{H}}_{tml}\right]$ and $\left[\hat{Q}_0,\hat{{H}}_{tml}\right]$. The remaining commutators, which do not involve the term $\hat{{H}}_{tml}$, are evaluated using the rules for the functions of operators.  The Heisenberg equations must be solved with the initial condition that, at the initial time, any operator in the Heisenberg picture should be equal to the corresponding operator in the Schr\"{o}dinger picture, denoted as $\hat{O}^{(S)}$. Furthermore, we impose the boundary condition \ref{eq:Phi1} between the operators, specifically $\hat\phi(t;x=0)=\hat\Phi_0(t)$. The system of equations \ref{eq:Heis_1}-\ref{eq:tlq} are the quantized version of the system of equations \ref{eq:Ham1}-\ref{eq:Ham6}.

Combining Eqs. \ref{eq:Heis_1} and \ref{eq:Heis_2}, applying the relation $[\bold{C}_b]\mathbf{p}=\mathbf{1}_N $ and using Eq. \ref{eq:Heis_6}, we obtain the Heisenberg equation for $\hat{\mathbf{\Phi}}$
\begin{equation}
\label{eq:quantum1}
[\mathbf{C}_b]\frac{d^2\hat{\mathbf{\Phi}}}{d t^2} +\frac{\partial U_b}{\partial \hat{\mathbf{\Phi}}}+  C_{c}\frac{d^2}{dt^2}\left({\hat{\Phi}_1}-{\hat{\Phi}_0}\right)\mathbf{1}_N=\mathbf{0}.
\end{equation}
From equations \ref{eq:Heis_5}, \ref{eq:Heis_6} and \ref{eq:Heis_1} we obtain the Heisenberg equation for $\hat{\Phi}_0$
\begin{equation}
\label{eq:quantum2}
C_{c}\frac{d^2}{dt^2}\left({\hat{\Phi}_1}-{\hat{\Phi}_0}\right)+\frac{1}{\ell} \hat{\phi}_x(t;x=0)=0.
\end{equation}
Lastly, from equations \ref{eq:tlphi} and \ref{eq:tlq} we obtain the Heisenberg equation for $\hat{\phi}$
\begin{equation}
\label{eq:quantum3}
\frac{\partial^2 \hat{\phi}}{\partial t^2}-v_p^2\frac{\partial^2 \hat{\phi}}{\partial x^2}=0 \qquad \text{for} \qquad 0<x<\infty.
\end{equation}
The system of equations \ref{eq:quantum1}-\ref{eq:quantum3} are the quantized version of the system of equations \ref{eq:eqlagrPhicir1}-\ref{eq:eqlagrphitml1}. 

The Heisenberg equations align with the results obtained when representing the semi-infinite transmission line as a cascade of discrete LC lumped elements and taking the continuum limit, \cite{peropadre_scattering_2013}.

\section{Reduced quantum model}

Solving Eqs. \ref{eq:tlphi} and \ref{eq:tlq} with the initial conditions $\hat{\phi}(t=0;x)=\hat{\phi}^{(S)}(x)$, $\hat{q}(t=0;x)=\hat{q}^{(S)}(x)$ for $0<x<\infty$,
where $\hat{\phi}^{(S)}(x)$ and $\hat{q}^{(S)}(x)$ are the flux field operator and the charge density field operator in the Schr\"odinger picture, we can express the current intensity operator $\hat{I}_0(t)=-\hat{\phi}_x(t;x=0)/\ell$ as functions of the voltage operator $\hat{V}_0(t)=\dot{\hat{\Phi}}_0$. This approach, which is at the basis of the input-output theory (e.g., \cite{gardiner_c_w_quantum_2004}), is widely used in the field of circuit quantum electrodynamics (e.g., \cite{vool_introduction_2017}, \cite{blais_circuit_2021})). We obtain (see Appendix \ref{sec:Reduced})
\begin{equation}
\label{eq:char_line}
{\hat{V}}_0=Z_c \hat{I}_0 + \hat{e}_0,
\end{equation}
where $Z_c=\sqrt{\ell/c}$ is the characteristic impedance of the line, $\hat{e}_0(t)=2\hat{v}_0^{\leftarrow}(t)$ and $\hat{v}_0^{\leftarrow}$ is the backward voltage wave operator given by \ref{eq:v+0}, which takes into account the contribution of the initial conditions of the transmission line. 

By using equation \ref{eq:char_line} the system of Heisenberg's equations \ref{eq:Heis_1}-\ref{eq:tlq} is reduced to the system of equations,
\begin{subequations}
\label{eq:red0}
\begin{align}
\label{eq:Heis_10}
\dot{\hat{\mathbf{\Phi}}}=[\bold{C}]^{-1}{\hat{\bold{Q}}}+\mathbf{p}\hat{Q}_0, \\
\label{eq:Heis_11}
\dot{\hat{\mathbf{Q}}}=-\frac{\partial{U}_c}{\partial \hat{{\mathbf{\Phi}}}},\\
\label{eq:Heis_12}
\dot{\hat{Q}}_0+\frac{1}{\tau}{\hat{Q}}_0=-\frac{1}{Z_c} \mathbf{p}^\intercal \hat{\mathbf{Q}}+\frac{1}{Z_c} \hat{e}_0,
\end{align}
\end{subequations}
where $\tau=Z_c C'_c$. Equations \ref{eq:Heis_10}-\ref{eq:Heis_12} must be solved with the initial conditions $\hat{\mathbf{\Phi}}(t=0)=\hat{\mathbf{\Phi}}^{(S)}$, $\hat{\mathbf{Q}}(t=0)=\hat{\mathbf{Q}}^{(S)}$ and $\hat{{Q}}_0(t=0)=\hat{{Q}}_0^{(S)}$.  These equations are the Heisenberg's equations of the equivalent circuit shown in Fig. 2, which is widely used in the literature. 

By expressing the operator $\hat{Q}_0$ in terms of the voltage operator $\hat{V}_0$, the system \ref{eq:Heis_1}-\ref{eq:tlq} reduces to
\begin{subequations}
\label{eq:red}
\begin{align}
\label{eq:Heis_13}
\dot{\hat{\mathbf{\Phi}}}=[\bold{A}]{\hat{\bold{Q}}}+\mathbf{p}C'_c\hat{V}_0, \\
\label{eq:Heis_14}
\dot{\hat{\mathbf{Q}}}=-\frac{\partial{U}_c}{\partial \hat{{\mathbf{\Phi}}}},\\
\label{eq:Heis_15}
\dot{{\hat{V}}}_0+\frac{1}{\tau}\hat{V}_0 = \mathbf{p}^\intercal \dot{\mathbf{Q}}(t)+\frac{1}{\tau}\hat{e}_0
\end{align}
\end{subequations}
where $[\bold{A}]=[\bold{C_b}]^{-1}-C_p\mathbf{p}\mathbf{p}^\intercal$. Equation \ref{eq:Heis_15} must be solved with the initial condition $\hat{V}_0(t=0)=\hat{V}_0^{(S)}$ where $\hat{V}_0^{(S)}=\mathbf{p}^\intercal \hat{\mathbf{Q}}^{(S)}+\frac{1}{C'_c}\hat{Q}_0^{(S)}$ is the voltage operator $\hat{V}_0$ in the  Schr\"{o}dinger picture.
The system of equations \ref{eq:red} can be put in the quantum Langevin form, as follows. 
From equation \ref{eq:Heis_15}, we obtain
\begin{equation}
  \hat{V}_0(t)=\mathbf{p}^\intercal g(t)*\dot{\hat{\mathbf{Q}}}+\frac{2}{\tau}g(t)*\hat{v}_{\leftarrow}^{(0)}(t)+g(t)\hat{V}_0^{(S)},
\end{equation}
where $g(t)=u(t)e^{-t/\tau}$ and the symbol $*$ denotes the time convolution product over the interval $[0, t ]$. Therefore, the system of equations \ref{eq:red0} can be further reduced to 
\begin{subequations}
\begin{align}
\dot{\hat{\mathbf{\Phi}}}=[\bold{A}]{\hat{\bold{Q}}}+[\bold{B}]g(t)*\dot{\hat{\mathbf{Q}}}+\hat{w}_{\leftarrow}^{(0)}(t), \\
\label{eq:Heis_14}
\dot{\hat{\mathbf{Q}}}=-\frac{\partial{U}_c}{\partial \hat{{\mathbf{\Phi}}}},
\end{align}
\end{subequations}
where $[\bold{B}]=C_p\mathbf{p}\mathbf{p}^\intercal$ and $\hat{w}_{\leftarrow}^{(0)}(t)=\frac{2}{\tau}C_p\mathbf{p}{g}(t)*\hat{v}_{\leftarrow}^{(0)}(t)+ C_p\mathbf{p}g(t)\hat{V}_0^{(S)}$.

\begin{figure}
    \centering
    \includegraphics[width=\columnwidth]{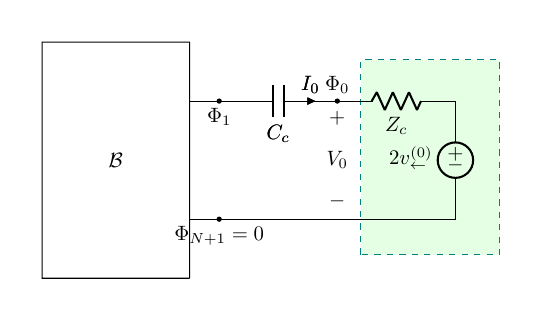}
    \caption{Equivalent lumped network of the semi-infinite transmission line capacitively coupled to the lumped circuit $\mathcal{B}$ shown in Fig. \ref{fig:SemiInfinite_TL_Network}.}
    \label{fig:Equivalent_SemiInf}
\end{figure}

\section{An application}

The reduced Heisenberg equations \ref{eq:red} (or \ref{eq:red0}) govern the time evolution of the observables of the lumped equivalent network shown in Fig. \ref{fig:Equivalent_SemiInf}. In the Heisenberg picture, the knowledge of the time evolution of the observables enables the evaluation of the statistical functions, such as expectation values, uncertainties, and correlation functions. 
However, finding solutions to Heisenberg equations may be a challenging task in general, and specifically when there are non-linear lumped elements. In this case, the numerical solution of the Heisenberg equations is needed  \cite{bender_solution_1983,razavy_heisenbergs_2011}.

\begin{figure}
    \centering\includegraphics[width=\columnwidth]{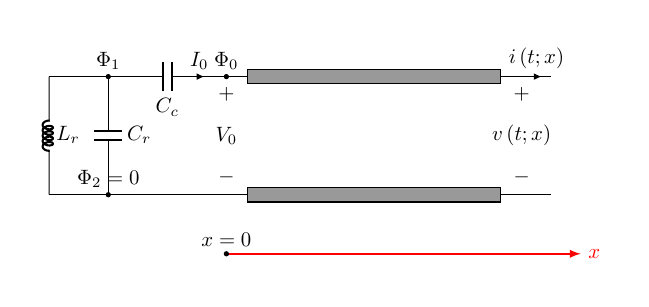}
    \caption{A parallel LC circuit capacitively coupled to a semi-infinite transmission line.}
    \label{fig:SemiInf_LC}
\end{figure}

In this section, we focus on a specific scenarios that admits an analytical solution due to the linearity. We investigate the network shown in Fig. \ref{fig:SemiInf_LC}. It consists of a semi-infinite transmission line capacitively coupled to the lumped circuit $\mathcal{B}$, consisting of a capacitor with capacitance $C_r$ connected in parallel to a linear inductor with inductance $L_r$.
By eliminating the observable $\hat{{Q}}_1$, the system of Heisenberg equations \ref{eq:red} simplifies to:
\begin{subequations}
\begin{align}
\label{eq:eq_51}
\ddot{\hat{{\Phi}}}_1+\frac{1}{L_r(C_r+C_c)} {\hat{\Phi}_1} =\frac{C_p}{C_r}\dot{\hat{V}}_0, \\
\label{eq:eq_52}
\dot{{\hat{V}}}_0+\frac{1}{\tau}\hat{V}_0=-\frac{1}{L_r C_r}\hat{\Phi}_1+\frac{2}{\tau}\hat{v}^{(0)}_{\leftarrow},
\end{align}
\end{subequations}
where $C_p=C_cC_r/(C_c+C_r)$ and $\tau=Z_c C_p$. The expression of the operator $\hat{V}_0$ in the Schr\"{o}dinger picture is given by $\hat{V}_0^{(S)}=\frac{1}{C_r}\hat{{Q}_1}^{(S)}+\frac{1}{C_p}\hat{Q}_0^{(S)}$.

When $1/\tau$ is significantly higher than the highest characteristic frequency of the system $\Omega_c$ we can neglect the first term on the left-hand side of equation \ref{eq:eq_52}. Thus, by combining the two equations, we obtain:
\begin{equation}
\label{eq:Heis_21}
\ddot{\hat{{\Phi}}}_1+\omega_r \alpha g^2 \dot{\hat{\Phi}}_1+\Omega_r^2\hat{\Phi}_1\cong2 g \dot{\hat{v}}^{(0)}_{\leftarrow}(t),
\end{equation}
where $\omega_r=1/\sqrt{L_rC_r}$, $g=C_c/(C_r+C_c)$, $\alpha=Z_c/Z_r$ and $Z_r=\sqrt{L_r/C_r}$. $\Omega_r=\omega_r\sqrt{1-g}$ is the renormalized resonance frequency of the LC circuit and $\kappa=\omega_r \alpha g^2$ is the decay rate of the oscillation amplitude due to the coupling with the line, in the so-called weakly coupling limit.
Equation \ref{eq:Heis_21} is an equation of the Heisenberg-Langevin type in the Markovian approximation (e.g., \cite{gardiner_c_w_quantum_2004}). This result aligns with those found in the literature, obtained in the weakly coupling limit using approximate effective Hamiltonians (e.g., \cite{peropadre_scattering_2013}, \cite{blais_circuit_2021}). 

If the term $\dot{{\hat{V}}}_0$ in equation \ref{eq:eq_52} cannot be ignored, we need to solve the complete system of equations \ref{eq:eq_51} and \ref{eq:eq_52}.
To address this, we utilize Laplace transforms, resulting in
\begin{equation}
\label{eq:rel_lapl_semi}
\left[
\begin{array}{c}
     \hat{\Phi}_1^{\mathtt{L}}(s)  \\
      \hat{V}_0^{\mathtt{L}}(s)
\end{array}
\right]
= 
\mathbf{H}(s)
\left[
\begin{array}{c}
\hat{F}_1 (s)  \\
\hat{F}_2 (s)
\end{array}
\right],
\end{equation}
where $\hat{\Phi}_1^{\mathtt{L}}(s)$ and $\hat{V}_0^{\mathtt{L}}(s)$ are, respectively, the Laplace transforms of $\hat{\Phi}_1 (t)$ and $\hat{V}_0 (t)$; $\mathbf{H}(s)$ is the classical transfer function matrix of the network whose entries are given by
\begin{subequations}
    \begin{align}
        (\mathbf{H})_{11}(s) &= \frac{1}{\omega_r^2} \frac{1}{ p(s)} \left[ \alpha g \left( \frac{s}{\omega_r} \right) + 1 \right], \\
        (\mathbf{H})_{12}(s) &= \frac{\alpha g}{\omega_r} \frac{1}{ p(s)} \left( \frac{s}{\omega_r} \right),   \\
        (\mathbf{H})_{21}(s) &= -\frac{\alpha g}{\omega_r} \frac{1}{ p(s)},  \\
        (\mathbf{H})_{22}(s) &= \frac{1}{ p(s)}  \left[ \left(\frac{s}{\omega_r}\right)^2 + (1 - g)\right];
    \end{align}
\end{subequations}
$ \hat{F}_1(s)$ and $\hat{F}_2(s)$ are known operators given by
\begin{subequations}
    \begin{align}
        \hat{F}_1 (s) &= s \hat{\Phi}_1^{(S)} + \frac{1}{C_r} \left( \hat{Q}_1^{(S)} + \hat{Q}_0^{(S)}\right) + \frac{C_p}{C_r}  \hat{V}_0^{(S)},   \\
       \hat{F}_2 (s) &= \tau \hat{V}_0^{(S)} + 2\hat{v}^{\mathtt{(0)L}}_{\leftarrow}(s),
    \end{align}
\end{subequations}
and $\hat{v}^{\mathtt{(0)L}}_{\leftarrow}(s)$ is the Laplace transform of $\hat{v}^{\mathtt{(0)}}_{\leftarrow}(t)$; $p(s)$ is the determinant of $\mathbf{H}^{-1}(s)$,
\begin{equation}
\label{eq:algebric}
p(s)=\alpha g \left(\frac{s}{\omega_r}\right)^3 + \left(\frac{s}{\omega_r}\right)^2 + \alpha g \left(\frac{s}{\omega_r}\right)+(1-g).
\end{equation}
The parameter $g=C_c/(C_r+C_c)$ varies in the interval (0,1): for $g=0$ the LC circuit is decoupled from the semi-infinite transmission line, for $g=1$ the system behaves as if the LC circuit were connected to the line via a short circuit, effectively creating an RLC circuit. The parameter $\alpha=Z_c/Z_r$ varies in the interval $(0,\infty)$.
We note that the entries $(\mathbf{H})_{11}$, $(\mathbf{H})_{12}$ and $(\mathbf{H})_{21}$ are dependent on both $s/\omega_r$ and $\omega_r$, whereas $(\mathbf{H})_{22}$ solely depends on $s/\omega_r$.  Since $\hat{V}_0^{(S)}=\frac{1}{C_r}\hat{{Q}_1}^{(S)}+\frac{1}{C_p}\hat{Q}_0^{(S)}$, both $\hat{F}_1$ and $\hat{F}_2$ are linear combinations of the operators $\hat{\Phi}_1^{(S)}$, $\hat{Q}_1^{(S)}$, $\hat{Q}_0^{(S)}$ and $\hat{v}^{\mathtt{(0)L}}_{\leftarrow}$.
\begin{figure*}
    \centering
    \includegraphics[width=0.7\textwidth]{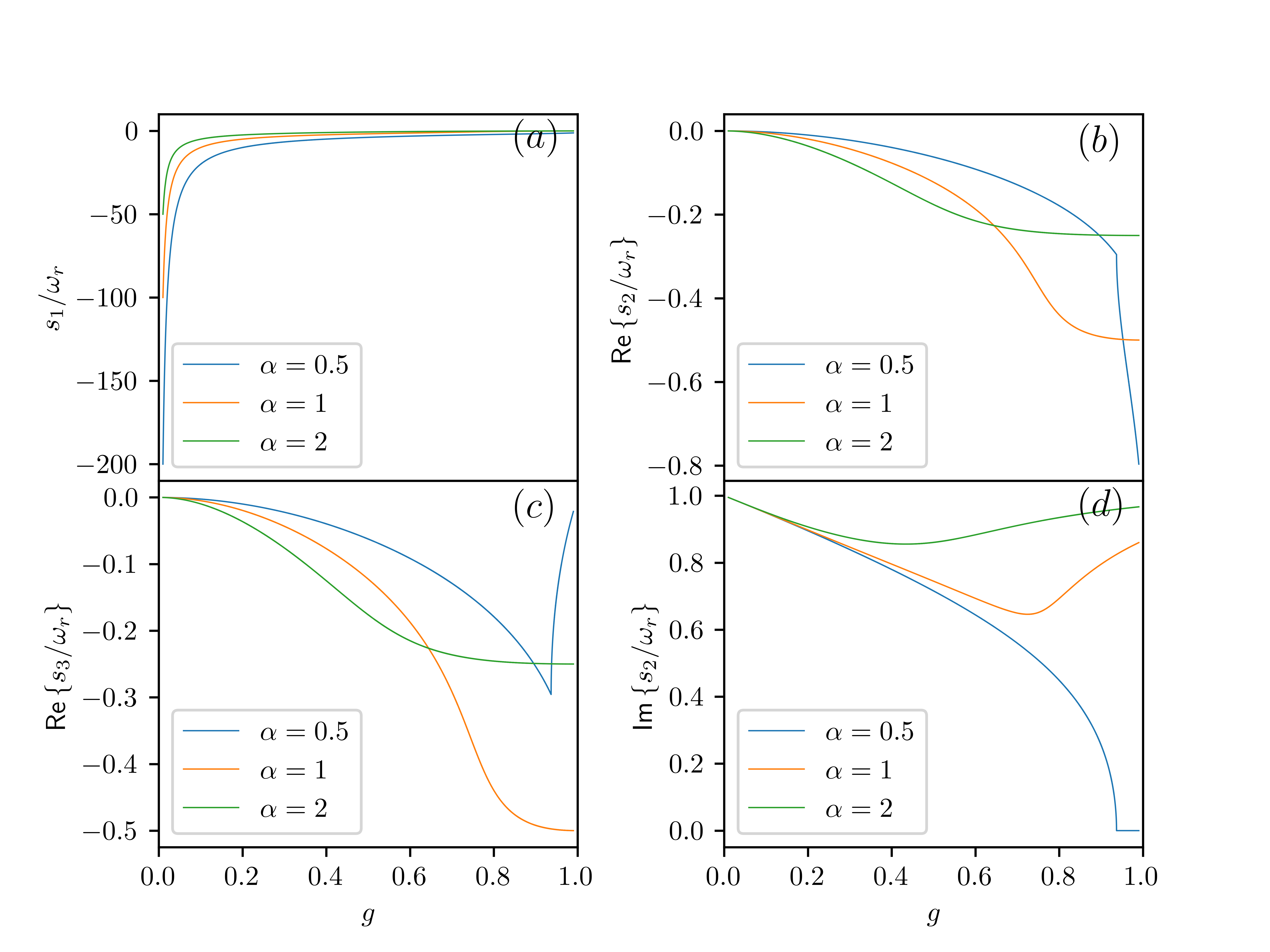}
    \caption{Behaviour of the poles, normalized to $\omega_r$, of the transfer matrix $\mathbf{H}(s)$ of a semi-infinite transmission line coupled capacitively to an LC circuit as the parameter $g$ varies, for $\alpha=0.5, 1.0, 2.0$.}
    \label{fig:Poles_SemiInf}
\end{figure*}

The poles of the matrix $\mathbf{H}$ correspond to the zeros of the polynomial $p(s)$ and are located to the left of the imaginary axis in the complex plane. We denote them by ${s}_1$, ${s}_2$ and ${s}_3$, where ${s}_1$ is real, and the other two may be complex conjugate. Consequently, the equivalent reduced network exhibits three natural modes that decay exponentially: in fact, the LC circuit coupled to a semi-infinite transmission line behaves as an open system. One mode is aperiodic and the other two modes can oscillate. The decay rate of the aperiodic mode is $-s_1$, the remaining two modes can oscillate with the natural frequency $\mbox{Im}\{s_2\}$. Figure \ref{fig:Poles_SemiInf} shows the behaviour of the normalized poles ${s}_1/\omega_r$, ${s}_2/\omega_r$ and ${s}_3/\omega_r$ as $g$ varies for $\alpha=0.5, 1.0, 2.0$. 

The normalized pole ${s}_1/\omega_r$ diverges as $g$ tends to $0$ (that is, as $C_c$ tends to $0$), and gradually approaches $0$ as $g$ tends to $1$ (that is, as $C_c$ tends to $\infty$), as shown in Fig. \ref{fig:Poles_SemiInf}(a). We now consider the normalized poles ${s}_2/\omega_r$ and ${s}_3/\omega_r$.  In the weak coupling limit $\alpha g \ll 1$ (implying  $\omega_r\tau \ll 1 $), the approximate equation $\ref{eq:Heis_21}$ indicates the presence of two oscillating modes with a normalized decay rate $\kappa/\omega_r=-\mbox{Re}\{{s}_2/\omega_r\}=-\mbox{Re}{\{{s}_3/\omega_r\}}\cong \alpha g^2$ (Figs. \ref{fig:Poles_SemiInf} (b) and \ref{fig:Poles_SemiInf} (c)) and with normalized oscillating frequency $\Omega_r/\omega_r=\mbox{Im}\{{s}_2/\omega_r\}\cong \sqrt{1-g}$ (Fig. \ref{fig:Poles_SemiInf}d).
As the coupling parameter $\alpha g$ deviates from being small, the behavior of $s_2/\omega_r$ and ${s_3/\omega_r}$ becomes complicated as $g$ and $\alpha$ vary. For $\alpha=1$ and $\alpha=2$ the normalized decay rate of the oscillating modes first increases and then saturates to values depending on $\alpha$ as $g$ tends to $1$ (Fig. \ref{fig:Poles_SemiInf}b and 6c); the normalized oscillating frequency initially decreases and then increases, although it remains lower than $1$. As $g$ approaches 1, the system behaves as a parallel RLC circuit. In this limit, the two oscillating modes transition to aperiodic modes for $\alpha\le0.5$ (as in a parallel RLC circuit): the imaginary part of ${s}_2/\omega_r$ vanishes for $g$ greater than a certain value, which depends on $\alpha$ as shown in Fig. \ref{fig:Poles_SemiInf}d. Finally, we also observe that when $\alpha$ tends to zero $\mbox{Im}\{{s}_2/\omega_r\}\cong \sqrt{1-g}$, while when $\alpha$ tends to infinity $\mbox{Im}\{{s}_2/\omega_r\}\cong 1$. In the limit $g \ll 1$, the decay rate of the oscillating modes increases linearly as $\alpha$ increases; instead, for values of $g$ of the order of one, the decay rate decreases as $\alpha$ increases, as shown in Figs. 6b and 6c (the curves cross between them for values of $g$ greater than $0.6$).

\begin{figure}
    \centering
    \includegraphics[width=0.5\textwidth]{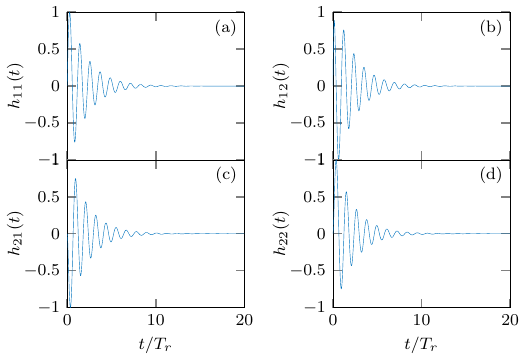}
    \caption{Entries of the impulse response matrix ${\mathbf{h}}(t)$ of a semi-infinite transmission line, capacitively coupled to an LC circuit, with $g=0.3$ and $\alpha=2$ where $T_r=2\pi/\omega_r$; each impulse response is normalized to its maximum absolute value.}
    \label{fig:Impulse_SemiInf03}
\end{figure}

\begin{figure}
    \centering
    \includegraphics[width=0.5\textwidth]{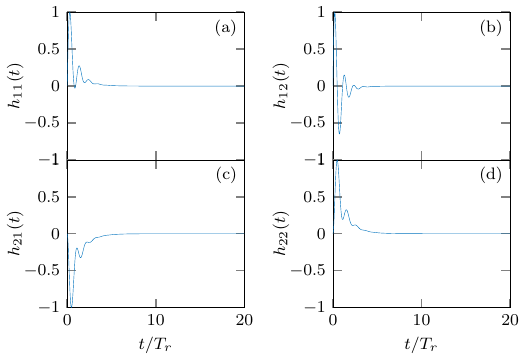}
    \caption{Entries of the impulse response matrix ${\mathbf{h}}(t)$ of a semi-infinite transmission line, capacitively coupled to an LC circuit, with $g=0.8$ and $\alpha=2$ where $T_r=2\pi/\omega_r$; each impulse response is normalized to its maximum absolute value}
    \label{fig:Impulse_SemiInf08}
\end{figure}

In the time domain, the relation \ref{eq:rel_lapl_semi} becomes
\begin{equation}
\left[
\begin{array}{c}
     \hat{\Phi}_1(t)  \\
      \hat{V}_0(t)
\end{array}
\right]
= 
\mathbf{h}(t)*
\left[
\begin{array}{c}
\hat{f}_1 (t)  \\
\hat{f}_2 (t)
\end{array}
\right],
\end{equation}
where $\mathbf{h}(t)$, $\hat{f}_1 (t)$ and $\hat{f}_2 (t)$ are, respectively, the inverse Laplace transform of $\mathbf{H}(s)$, $\hat{F}_1 (s)$ and $\hat{F}_2 (s)$, and the symbol $*$ denotes the time convolution product over the interval $[0, t ]$. The expressions of the operators $\hat{f}_1 (t)$ and $\hat{f}_2 (t)$ are
\begin{subequations}
    \begin{align}
        \hat{f}_1 &= \hat{\Phi}_1^{(S)}\dot{\delta}(t) + \left[\frac{1}{C_r} \left( \hat{Q}_1^{(S)} + \hat{Q}_0^{(S)}\right) + \frac{C_p}{C_r}\hat{V}_0^{(S)}\right]\delta(t),   \\
       \hat{f}_2 &= \tau \hat{V}_0^{(S)}\delta(t) + 2\hat{v}^{\mathtt{(0)}}_{\leftarrow}(t).
    \end{align}
\end{subequations}
The matrix $\mathbf{h}(t)$ represents the classical impulse response of the network. Specifically, the elements $h_{11}(t)$ and $h_{21}(t)$ correspond to the responses of the flux ${\Phi}_1(t)$ and the voltage $V_0(t)$ when the inductor is instantaneously charged by an impulsive voltage source with unitary amplitude connected in series with the inductor. On the other hand, $h_{12}(t)$ and $h_{22}(t)$ represent the responses of ${\Phi}_1(t)$ and $V_0(t)$ when the backward voltage wave at $x=0$ is a Dirac pulse with amplitude equal to $0.5$.

The entries of the impulse response matrix $\mathbf{h}(t)$ are numerically evaluated by using the IFFT algorithm. In Fig. \ref{fig:Impulse_SemiInf03}, we show the impulse response matrix $\mathbf{h}(t)$ for $g=0.3$ and $\alpha=2$ where $T_r=2\pi/\omega_r$. Each entry of $\mathbf{h}(t)$ is normalized to its maximum absolute value. For these specific values of  $\alpha$ and $g$ the decay rate of the aperiodic mode is roughly one order of magnitude higher than the decay rate of the oscillating modes. Consequently, in this scenario, the contribution of the damped oscillating modes dominates. In Fig. \ref{fig:Impulse_SemiInf08}, we show the normalized entries of $\mathbf{h}(t)$ for $g=0.8$ and $\alpha=2$. In this case, the contribution of the aperiodic mode becomes evident as its decay rate is lower than the decay rate of the oscillating modes.

\section{Conclusions}

The quantization of systems composed of transmission lines connected to lumped circuits is challenging due to the coexistence of continuous and discrete degrees of freedom. Yurke and Denker proposed a direct quantization scheme for these systems, constructing the Hamiltonian directly in terms of fields. Specifically, the lumped circuit contributions at the transmission line ends are represented using Lagrangian densities incorporating Dirac $\delta$-functions. However, this approach may lead to some issues.  For instance, when dealing with capacitive coupling with flux variables, the momentum densities are distributions with Dirac $\delta$-functions. Consequently, there may be the need to handle products of $\delta$-functions, which requires regularization procedures.  In this paper, we introduce a $\delta$-free Lagrangian formulation without the need for transmission line discretization or mode expansions. Upon solving the Heisenberg equations for line observable we obtain the equivalent one-port model commonly used in the literature for semi-infinite transmission lines.
The proposed approach readily generalizes to more complex coupling configurations. This encompasses scenarios where the transmission line is connected to the lumped circuit through various combinations of capacitors and inductors, both linear and nonlinear. Furthermore, it is applicable to broader situations, including finite-lenght transmission lines and multiconductor transmission lines. 

\onecolumngrid

\appendix

\section{Evaluation of the commutators that involve $\hat{{H}}_{tml}$}
\label{sec:Heisen}
In this appendix, we evaluate the commutators $\left[\hat{\phi},\hat{{H}}_{tml}\right]$,  $\left[\hat{q},\hat{{H}}_{tml}\right]$ and $\left[\hat{Q}_0,\hat{{H}}_{tml}\right]$.

\subsection{Evaluation of $\left[\hat{\phi}(t;x),\hat{{H}}_{tml}\right]$ and $\left[\hat{q}(t;x),\hat{{H}}_{tml}\right]$ for $0<x<d$}
\label{sec:Commutator}

We first consider the commutator $\left[\hat{\phi}(t;x),\hat{{H}}_{tml}\right]$. We have the following:
\begin{equation}
\left[\hat{\phi}(t;x),\hat{{H}}_{tml}\right] =\frac{1}{2c}\int_0^{\infty} dx' \left[\hat{\phi}(t;x),\hat{q}^2(t;x')\right]=\frac{1}{c}\int_0^{\infty} dx' \left[\hat{\phi}(t;x),\hat{q}(t;x')\right]\hat{q}(t;x').
\end{equation}
Therefore, by applying the commutation relation \ref{eq:com3} we obtain
\begin{equation}
\left[\hat{\phi}(t;x),\hat{{H}}_{tml}\right]=\frac{i\hbar}{c}\hat{q}(t;x) \qquad \text{for} \qquad 0<x<{\infty}.
\end{equation}

Now we consider the commutator $\left[\hat{q}(t;x),\hat{{H}}_{tml}\right]$. We have the following:
\begin{equation}
\label{eq:com_tml_1}
\left[\hat{q}(t;x),\hat{{H}}_{tml}\right] =\frac{1}{2\ell}\int_0^{\infty} dx' \left[\hat{q}(t;x),\hat{\phi}_x^2(t;x')\right]=\frac{1}{\ell}\int_0^{\infty} dx' \left[\hat{q}(t;x),\hat{\phi}_x(t;x')\right]\hat{\phi}_x(t;x'),
\end{equation}
where $\hat{\phi}_{x}$ denotes the first partial derivative with respect to the variable $x$. From the commutation relation \ref{eq:com3} we obtain, for $0<x<{\infty}$,
\begin{equation}
\label{eq:com_tml_2}
\left[\hat{q}(t;x),\hat{\phi}_x(t;x')\right]=-i\hbar\frac{\partial}{\partial x'}\delta(x'-x).
\end{equation}
Combining the relations \ref{eq:com_tml_1} and \ref{eq:com_tml_2}, and using integration by parts, we eventually have
\begin{equation}
\left[\hat{q}(t;x),\hat{{H}}_{tml}\right]=\frac{i\hbar}{\ell}\hat{\phi}_{xx}(t;x) \qquad \text{for} \qquad 0<x<{\infty},
\end{equation}
where $\phi_{xx}$ denotes the second partial derivative with respect to the variable $x$.

\subsection{Evaluation of $\left[\hat{Q}_0,\hat{{H}}_{tml}\right]$}

We now consider the commutators $\left[\hat{Q}_0,\hat{{H}}_{tml}\right]$. We have the following
\begin{equation}
\label{eq:com_tml_3}
\left[\hat{Q}_0,\hat{{H}}_{tml}\right] =\frac{1}{2\ell} \left[\hat{Q}_0,\int_0^{\infty} dx'\hat{\phi}_x^2(t;x')\right]=\frac{1}{2\ell} \left[\hat{Q}_0,\int_0^{\Delta x} dx'\hat{\phi}_x^2(t;x')\right]
\end{equation}
where $\Delta x$ is an arbitrarily small interval. Since
\begin{equation}
\int_0^{\Delta x} dx'\hat{\phi}_x^2(t;x')=\Delta x \left[\frac{\hat{\phi}(t;x=\Delta x)-\hat{\phi}(t;x=0)}{\Delta x}\right]^2+\mathcal{O}(\Delta x)
\end{equation}
we obtain
\begin{equation}
\left[\hat{Q}_0,\hat{{H}}_{tml}\right] =\frac{1}{2\ell}\frac{1}{\Delta x} \left[\hat{Q}_0,\left({\hat{\phi}(t;x=\Delta x)-\hat{\phi}(t;x=0)}\right)^2\right]+\mathcal{O}(\Delta x).
\end{equation}
Recalling the boundary condition \ref{eq:Phi1} , i.e. $\hat{\phi}(t;x=0)=\hat{\Phi}_0$ , and using \ref{eq:com2}, we have
\begin{equation}
\left[\hat{Q}_0,\hat{{H}}_{tml}\right] =\frac{\hbar}{\ell}\frac{{\hat{\phi}(t;x=\Delta x)-\hat{\phi}(t;x=0)}}{\Delta x} +\mathcal{O}(\Delta x).
\end{equation}
By taking now the limit $\Delta x \rightarrow 0$ we obtain
\begin{equation}
\left[\hat{Q}_0,\hat{{H}}_{tml}\right]=\frac{i\hbar}{l}\hat{\phi}_{x}(t;x=0).
\end{equation}


\section{Solution of the transmission line equation \ref{eq:quantum3}}
\label{sec:Reduced}

Solving Eqs. \ref{eq:tlphi} and \ref{eq:tlq} with the initial conditions for $0<x<\infty$
\begin{equation}
\label{eq:initconflux}
\hat{\phi}(t=0;x)=\hat{\phi}^{(S)}(x),
\end{equation}
\begin{equation}
\label{eq:initconchar}
\hat{q}(t=0;x)=\hat{q}^{(S)}(x),
\end{equation}
where $\hat{\phi}^{(S)}(x)$ and $\hat{q}^{(S)}(x)$ are the flux field operator and the charge density field operator in the Schr\"odinger picture, we can express the current intensity operator $\hat{I}(t)=-\hat{\phi}_x(t;x=0)/\ell$ as functions of the transmission line voltage operator $\hat{V}(t)=\dot{\hat{\Phi}}_0$ at $x=0$.

The general solution of \ref{eq:quantum3} is for $0<x<\infty$ and $t>0$
\begin{equation}
\label{eq:phisol0}
\hat{\phi}(t;x)=\hat{\phi}_{\rightarrow}\left(t-\frac{x}{v_p}\right)+\hat{\phi}_{\leftarrow}\left(t+\frac{x}{v_p}\right),
\end{equation}
where $v_p=1/\sqrt{\ell c}$ is the propagation velocity of the electromagnetic signal along the line. The operators $\hat{\phi}_{\rightarrow}=\hat{\phi}_{\rightarrow}(\eta_-)$ and $\hat{\phi}_{\leftarrow}=\hat{\phi}_{\leftarrow}(\eta_+)$ are unknown: they are defined, respectively, in the interval $(-\infty,+\infty)$ and $(0,+\infty)$. They depend on the initial conditions \ref{eq:initconflux}, \ref{eq:initconchar}, and the interaction with the circuit $\mathcal{B}$. The expression of the charge density field operator is for $0<x<\infty$ and $t>0$
\begin{equation}
\label{eq:qsol}
\hat{q}(t;x)={c}\left[{\dot{\hat{\phi}}}_{\rightarrow} \left(t-\frac{x}{v_p}\right)+{\dot{\hat{\phi}}}_{\leftarrow}\left(t+\frac{x}{v_p}\right)\right],
\end{equation}
where the point on the top denotes the first ordinary derivative with respect to time. 

The initial conditions \ref{eq:initconflux}, \ref{eq:initconchar} for the flux and charge density field operators determine the forward and backward flux operators $\hat{\phi}_{\rightarrow}$ and $\hat{\phi}_{\leftarrow}$ in the interval $(-\infty,0)$ and $(0,\infty)$, respectively. We have for $0<x<\infty$
\begin{equation}
\label{eq:phisol0init}
\hat{\phi}^{(S)}(x)=\hat{\phi}_{\rightarrow}\left(-\frac{x}{v_p}\right)+\hat{\phi}_{\leftarrow}\left(\frac{x}{v_p}\right),
\end{equation}
\begin{equation}
\label{eq:qsol0init}
\hat{q}^{(S)}(x)={c}\left[{\dot{\hat{\phi}}}_{\rightarrow}\left(-\frac{x}{v_p}\right) + {\dot{\hat{\phi}}}_{\leftarrow}\left(\frac{x}{v_p}\right)\right].
\end{equation}
From these equations, we obtain for $0<x<\infty$
\begin{equation}
\label{eq:phisol1}
{\dot{\hat{\phi}}}_{\rightarrow}\left(-\frac{x}{v_p}\right)= \frac{1}{2c}\left[\hat{q}^{(S)}(x)- \frac{1}{Z_c}{\hat{\phi}^{(S)}_x(x)}\right],
\end{equation}
\begin{equation}
\label{eq:qsol1}
{\dot{\hat{\phi}}}_{\leftarrow}\left(\frac{x}{v_p}\right)= \frac{1}{2c}\left[\hat{q}^{(S)}(x)+ \frac{1}{Z_c}\hat{\phi}_x^{(S)}(x)\right].
\end{equation}

We now introduce the voltage and current intensity field operators given by $\hat{v}=\hat{q}/c$ and $\hat{i}=-\hat{\phi}_x/\ell$. From \ref{eq:phisol0} and \ref{eq:qsol} we obtain
\begin{equation}
\label{eq:vsol}
\hat{v}(t;x)=\hat{v}_{\rightarrow}\left(t-\frac{x}{v_p}\right)+\hat{v}_{\leftarrow}\left(t+\frac{x}{v_p}\right),
\end{equation}
\begin{equation}
\label{eq:isol}
\hat{i}(t;x)=\frac{1}{Z_c}\left[\hat{v}_{\rightarrow}\left(t-\frac{x}{v_p}\right)-\hat{v}_{\leftarrow}\left(t+\frac{x}{v_p}\right)\right],
\end{equation}
where (to simplify the notation) $\hat{v}_{\rightarrow}$ stays for $\dot{\hat{\phi}}_{\rightarrow}$ and $\hat{v}_{\leftarrow}$ stays for $\dot{\hat{\phi}}_{\leftarrow}$; $Z_c=\sqrt{\ell/c}$ is the characteristic impedance of the line. The field operators $\hat{v}_{\rightarrow}(t)$ and $\hat{v}_{\leftarrow}(t)$ are the forward and backward voltage wave operators of the line at $x=0$. The voltage and current distributions along the line are completely
determined by the functions $\hat{v}_{\rightarrow}$ and $\hat{v}_{\leftarrow}$, and vice versa.

We now can characterize the semi-infinite transmission line as a one-port. Specifying expressions \ref{eq:vsol} and \ref{eq:isol} at the line end $x=0$ we obtain
\begin{equation}
\label{eq:v1}
\hat{V}(t)=\hat{v}_{\rightarrow}\left(t\right)+\hat{v}_{\leftarrow}\left(t\right),
\end{equation}
\begin{equation}
\label{eq:i1}
Z_c\hat{I}(t)=\hat{v}_{\rightarrow}\left(t\right)-\hat{v}_{\leftarrow}\left(t\right).
\end{equation}
Subtracting Eqs. \ref{eq:v1} and \ref{eq:i1} termwise, we have for any $0\leq t <\infty$:
\begin{equation}
\label{eq:rel1}
\hat{V}(t)-Z_c\hat{I}(t)=2\hat{v}^0_{\leftarrow}(t).
\end{equation}
This is the equation that governs the time evolution of the one-port representing the transmission line, where the backward voltage operator
\begin{equation}
\label{eq:v+0}
{\hat{v}}_{\leftarrow}^0\left(t\right)= \frac{1}{2}\hat{q}^{(S)}(v_pt)+\frac{1}{2}v_p\hat{\phi}^{(S)}_x(v_pt)
\end{equation}
is known. The forward voltage wave operator for $-\infty<\eta_-<0$ depends only on the initial conditions and is given by
\begin{equation}
\label{eq:v_0}
\hat{v}_{\rightarrow}^0\left(\eta_-\right)= \frac{1}{2}\hat{q}^{(S)}[v_p(-\eta_-)]-\frac{1}{2}v_p \hat{\phi}_x^{(S)}[v_p(-\eta_-)].
\end{equation}

\bibliographystyle{ieeetr}

\end{document}